%% file: karmma_tomo_prd.tex
\def\jcap{JCAP}
\def\mnras{MNRAS}             
\def\aap{A\&A}             
\def\apjs{ApJS}             
\def\apjl{ApJL}             

\documentclass[%
reprint,
superscriptaddress,
nofootinbib,
nolongbibliography,
 amsmath,amssymb,
 aps,
prd,
floatfix,
]{revtex4-2}

\usepackage{graphicx}
\usepackage{dcolumn}
\usepackage{bm}
\usepackage{xcolor}

\definecolor{maroon}{RGB}{192,0,0}
\usepackage[colorlinks=true, citecolor=maroon,
linkcolor=maroon,
urlcolor=blue]{hyperref}
\usepackage{times}

\newcommand{\eps}{\epsilon}
\newcommand{\mmat}[1]{{\mathbf{#1}}}

\newcommand{\mvec}[1]{{\bm{#1}}}
\newcommand{\karmma}{{\sc karmma}~}

\begin{document}

\title{Bayesian mass mapping with weak lensing data using {\sc karmma} -- validation with simulations and application to Dark Energy Survey Year 3 data}

\author{Supranta S. Boruah}
 \email{supranta@sas.upenn.edu}
\affiliation{Department of Astronomy and Steward Observatory, University of Arizona, 933 N Cherry Ave, Tucson, AZ 85719, USA}
\affiliation{Department of Physics and Astronomy, University of Pennsylvania, Philadelphia, PA 19104, USA}
\author{Pier Fiedorowicz}
\affiliation{Department of Physics, University of Arizona, 1118 E. Fourth Street, Tucson, AZ, 85721, USA}
\affiliation{Lawrence Livermore National Laboratory, Livermore, CA 94550, USA}
\author{Eduardo Rozo}
\affiliation{Department of Physics, University of Arizona, 1118 E. Fourth Street, Tucson, AZ, 85721, USA}

\date{\today}

\begin{abstract}
We update the field-level inference code \karmma\ to enable tomographic forward-modelling of shear maps.  Our code assumes a lognormal prior on the convergence field, and properly accounts for the cross-covariance in the lensing signal across tomographic source bins.  We use mock weak lensing data from $N$-body simulations to validate our mass-mapping forward model by comparing our posterior maps to the input convergence fields. We find that \karmma\ produces more accurate reconstructions than traditional mass-mapping algorithms.  Moreover, the \karmma\ posteriors reproduce all statistical properties of the input density field we tested --- one- and two-point functions, and the peak and void number counts --- with $\approx 10\%$ accuracy.  Our posteriors exhibit a small bias that increases with decreasing source redshift, but these biases are small compared to the statistical uncertainties of current (DES) cosmic shear surveys.  Finally, we apply \karmma to Dark Energy Survey Year 3 (DES-Y3) weak lensing data,  and verify that the two point shear correlation function $\xi_+$ is well fit by the correlation function of the reconstructed convergence field.  This is a non-trivial test that traditional mass mapping algorithms fail. The code is publicly available at \url{https://github.com/Supranta/KaRMMa.git}. \karmma DES-Y3 mass maps are publicly available at \url{https://zenodo.org/records/10672062}. 
\end{abstract}

\maketitle


\input{sections/intro}
\input{sections/data}
\input{sections/formalism}
\input{sections/sim_test}
\input{sections/results}
\input{sections/conclusion}

\begin{acknowledgments}
We thank Marco Gatti for providing the DES-Y3 shape catalogue, and for useful comment on the manuscript that helped improved the clarity of our presentation. The computation presented here was performed on the High Performance Computing (HPC) resources supported by the University of Arizona TRIF, UITS, and Research, Innovation, and Impact (RII) and maintained by the UArizona Research Technologies department. SSB was supported by the Department of Energy Cosmic Frontier program, grant DE-SC0020215, and NSF grant 2009401. ER's work is supported is supported by NSF grant 2009401.  ER also receives funding from DOE grant DE-SC0009913 and NSF grant 2206688.
\end{acknowledgments}

\appendix
\input{sections/appendix/ln_test}
\bibliography{karmma_tomo}

\end{document}

%% file: sections/intro.tex
\section{Introduction}\label{sec:intro}

Most standard analyses of cosmological data use 2-point summary statistics such as correlation functions and power spectra for extracting information from these data sets \citep[e.g, ][]{DESY3_3x2pt_2022, eBOSS_2021}. However, we lose valuable non-Gaussian information while compressing the full data set to 2-point summary statistics. Field-based inference, where one extracts information from the full field, is emerging as a powerful alternative for optimally extracting information from cosmological data sets across a broad variety of probes, including CMB lensing \citep{Millea2019, Millea2020, Millea2021}, integrated Sachs-Wolfe (ISW) effect \cite{Zhou2023a}, galaxy clustering \citep{Jasche2019, Porth2023}, peculiar velocity \citep{Boruah2022_PV, PrideauxGhee2023, Bayer2023} and weak lensing \citep{Boruah2022, Porqueres2022, Zhou2023}. 

In this paper, we focus on field-based inference of weak lensing data. Such an analysis requires an accurate forward model of the observed shear field. Once we have an accurate forward model for the weak lensing data, we can use it to reconstruct the unknown dark matter distribution of the Universe. Indeed, reconstructed mass maps from weak lensing surveys are crucial for extracting non-Gaussian information \citep[e.g,][]{Euclid2023}, for studying voids \cite{Kovacs2022}, and for cross-correlation studies \citep{Liu2015, Hojjati2017}. As such, they are an essential data product for Stage-IV weak lensing surveys. However, the reconstruction of the unseen dark matter distribution from weak lensing shear data constitutes a non-trivial inverse problem. While standard inversion methods such as Kaiser-Squires inversion \citep{Kaiser1993} are widely used, they suffer a variety of problems. For example, these methods lead to biased reconstruction in the presence of a survey mask. Furthermore, in the presence of shape noise the reconstructed mass maps do not have the expected statistical properties.  For instance, the two-point functions of the recovered maps are inconsistent with direct measurements of the two point shear correlation function (see Figure~\ref{fig:desy3_corr_cl_comparison}).

Forward modelled mass-map reconstruction naturally address these difficulties. For example, by forward-modelling the convergence field in the masked region, we get mass maps free from masking effects \citep{Mawdsley2020, Fiedorowicz2022}. Likewise, depending on the accuracy of the forward model, we are able to reconstruct mass maps with the correct 2-point and one-point function. 

A number of different weak lensing field-based analysis methods have been proposed in the literature, though none so far have been applied to data. Running approximate numerical structure formation simulations is likely to be the most accurate forward model. This is the approach followed by \cite{Porqueres2021, Porqueres2022}, who used Lagrangian Perturbation Theory (LPT) to forward model the cosmic shear data within the {\sc borg} Bayesian forward modeling framework \citep{Jasche2013, Jasche2019}. However, this approach is computationally expensive and analyses are often limited to small areas with a limited resolution.

Because of these difficulties, a number of alternative field-based inference methods have tried to directly model the observed 2-dimensional projected fields, thereby bypassing the need to generate a three-dimensional matter density field. For example, \cite{Alsing2016, Alsing2017} developed a Bayesian hierarchical model that simultaneously inferred the mass distribution and power spectrum of the weak lensing maps in tomographic bins. However, their method assumed Gaussianity of the field and only worked on flat sky. \cite{Loureiro2023, Sellentin2023} proposed a similar algorithm,  {\sc almanac}, which performs wide-field mass mapping on a sphere, but still assumes Gaussianity. Similarly, \cite{Kovacs2022} sampled mass maps with DES-Y3 data using the constrained realization approach \cite{Hoffman1991, Zaroubi1995, KodiRamanah2019} assuming Gaussianity of the convergence field.

Since the late-time density field is highly non-Gaussian, it is important to include the non-Gaussianity of the density/convergence field in the forward model. To that end, \cite{Fiedorowicz2022, Fiedorowicz2022a} introduced the {\sc karmma} algorithm. In \karmma, the convergence field is modeled as a lognormal random field, enabling us to by-pass the need to run expensive numerical simulations to reconstruct the matter density field, while simultaneously improving upon the Gaussian assumption made in other works (e.g. {\sc almanac}). More recently \cite{Zhou2023} introduced {\sc miko}, a flat-sky field-level inference code that also utilizes the same lognormal prior for the convergence field. The lognormal model has been shown to be a good approximation of the late-time non-Gaussian convergence field \citep{Xavier2016, Clerkin2017}. Indeed, in our earlier works we demonstrated that the \karmma\ posteriors from mock observations of simulated shear maps accurately reproduced a broad range of statistical properties of the corresponding input convergence fields.  However, the algorithm had a severe limitation: it was only able to forward model a single tomographic source bin at a time.  Here, we extend the {\sc karmma} algorithm using the methodology of \cite{Boruah2022} to enable tomographic mass mapping on a sphere while modeling the non-Gaussianity of the convergence field. We validate the updated code by running it on mock weak lensing data based on $N$-body simulations. We find that the mass maps reconstructed with \karmma have smaller errors than traditional mass mapping approaches.  Further, the statistical properties of the posterior maps are consistent with those from numerical simulations. After validating our method with mock simulations, we apply \karmma to the Dark Energy Survey (DES) Year 3 (Y3) weak lensing data. We make our maps publicly available at \url{https://zenodo.org/records/10672062}. To the best of our knowledge, these are the first publicly-available Bayesian mass maps reconstructed with weak lensing data on the curved sky.

This paper is structured as follows: in section \ref{sec:data}, we present the simulations and the weak lensing data used in this work. In section \ref{sec:formalism}, we present the main features of \karmma. Then in section \ref{sec:sim_test}, we validate \karmma with mock weak lensing simulations before presenting the Bayesian mass maps produced from DES-Y3 data in section \ref{sec:results}. Finally, we conclude in section \ref{sec:conclusion}. In Appendix \ref{app:ln_tests}, we show the results of our tests with lognormal mocks.

%% file: sections/data.tex
\section{Data}\label{sec:data}

\subsection{Mock maps from simulations}\label{ssec:sims}

In this section, we describe the mock weak lensing catalogues used in section \ref{sec:sim_test} to test the performance of our code. We use the publicly available simulations of \cite[][hereafter T17]{Takahashi2017}. T17 produced a suite of $108$ mock weak lensing catalogues by ray-tracing $N$-body simulations run with cosmological parameters $\Omega_m = 0.279$, $\Omega_b = 0.046$, $h = 0.7$, $\sigma_8 = 0.82$ and $n_s = 0.97$. Each mock catalogue consists of full-sky shear and convergence maps on redshift shells separated by $150~h^{-1}$ Mpc. We use {\sc healpix} maps at a resolution of $N_{\text{side}}=4096$ for this work. 

\begin{figure}
    \centering
    \includegraphics[width=\linewidth]{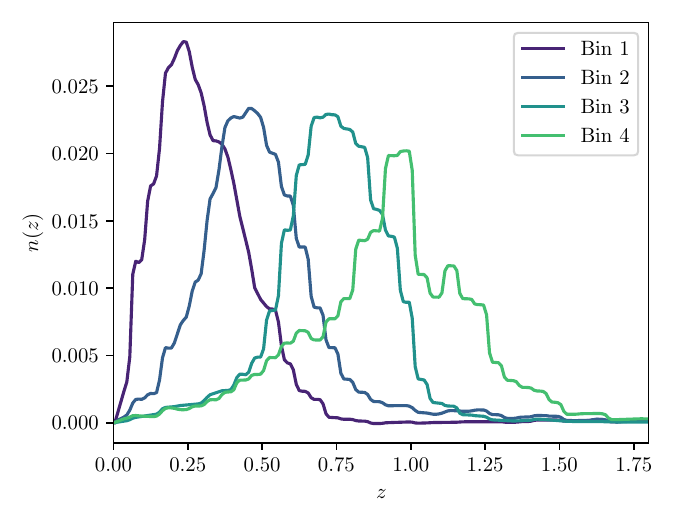}
    \caption{The source redshift distribution, $n(z)$, for the DES-Y3 used in this work.}
    \label{fig:desy3_nz}
\end{figure}

To generate simulated shear maps, we adopt the survey properties from the DES-Y3 analysis \cite{Amon2022, Secco2022}. Specifically, we use four tomographic redshift bins. The source redshift distribution of each bin is taken from \cite{Myles2021}, shown here in Figure \ref{fig:desy3_nz}. The effective number density in each of the tomographic bins is $[1.476,1.479,1.484,1.461]$ arcmin$^{-2}$. Finally, we adopt the shape noise estimate from the DES-Y3 shear catalog, $\sigma_{\epsilon} = 0.261$ \citep{DESY3_shapecatalogue}.

We produce tomographic maps of the quantity, $\mathcal{A}$, (here $\mathcal{A}$ can be either the convergence or shear map) by taking a weighted average of the maps at various redshift shells with a given redshift distribution, $n(z)$, via
\begin{equation}
    \mathcal{A}^{i} = \frac{\sum_{j=1}^{N_{\text{shells}}} n^i(z_j) \mathcal{A}_{\text{shell}}^{j}}{\sum_{j=1}^{N_{\text{shells}}} n^i(z_j)},
\end{equation}
where $n^{i}$ is the redshift distribution of the $i$-th tomographic bin and $\mathcal{A}^j_{\text{shell}}$ is the map of the quantity $\mathcal{A}$ on the $j$-th redshift shell. Note that the finite redshift resolution leads to a systematic error of $\mathcal{O}(\lesssim 5\%)$ on the power spectrum of the convergence field. Consequently, for the purposes of testing our code, we use the power spectrum measured from the T17 simulations, rather than a theory power spectrum computed using Boltzmann codes such as {\sc camb} \cite{camb} or {\sc class} \cite{class}. We change this in section \ref{sec:results} while applying to DES-Y3 data.

Our `observed' shear map is a noisy realization of the true shear. Specifically, we produce our noisy shear maps at a higher resolution of $N_{\rm side} = 1024$ and then downgrade to a resolution of $N_{\rm side}=256$ to mimic the impact of pixelization. The number of sources in each sky pixel is a Poisson random draw from the mean source density of the appropriate tomographic bin in the higher resolution maps. Note that this ignores source clustering which is an important systematic effect while considering non-Gaussian information \cite{Gatti2024}. We add to each shear component a Gaussian shape noise with zero mean and standard deviation, $\sigma_{\epsilon} / \sqrt{N_p}$, where, $N_p$ is the number of source galaxies sampled in that pixel. A resolution of $N_{\text{side}}=256$ corresponds to a pixel size with angular resolution of $\sim 13$ arcmin.

\subsection{DES-Y3 weak lensing  data}\label{ssec:desy3_data}

In section \ref{sec:results}, we run {\sc karmma} on the Dark Energy Survey Year 3 (DES-Y3) weak lensing data to produce Bayesian mass maps from weak lensing data. We use the DES-Y3 {\sc metacalibration} \cite{Huff2017, Sheldon2017} shape catalogue  \cite{DESY3_shapecatalogue}. These galaxies were divided into the $4$ tomographic bins according to their estimated photometric redshifts \cite{Myles2021}. Then we construct a {\sc healpix} map of estimated shear for each of the tomographic bin by taking the weighted average over the galaxy ellipticities, $\epsilon$, in each pixel as, 
\begin{equation}
    \gamma^{\alpha}_{p} = \frac{\sum_{i\in p} w_i \epsilon_{i,\alpha}}{\sum_{i\in p} w_i}, \alpha = 1,2,
\end{equation}
where $w_i$ are the per-galaxy inverse variance weights, $p$ is the pixel label, and $\alpha$ denotes the two polarizations. The estimated variance of each shear component in the pixel $p$ is given as \cite{Nicola2021, Doux2022}, 

\begin{equation}
    \sigma^2_{\epsilon,p} = \frac{\sum_{i \in p} w^2_i (e^{2}_{i,1} + e^{2}_{i,2})/2}{(\sum_{i \in p} w_i)^2}.
\end{equation}

%% file: sections/formalism.tex
\begin{figure*}
    \centering
    \includegraphics[width=\linewidth]{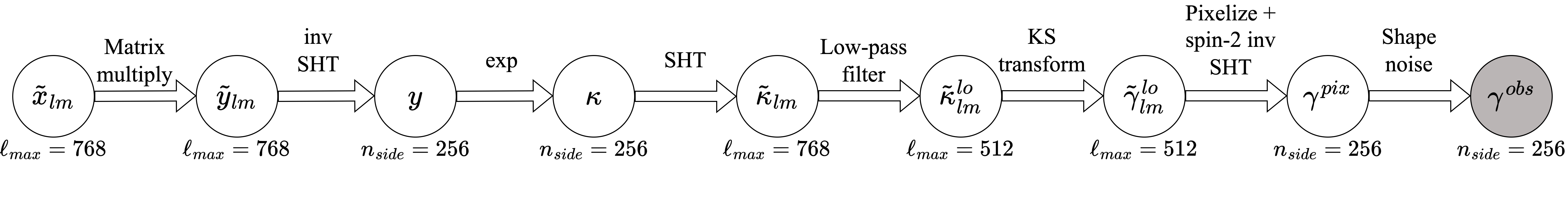}
    \caption{Illustration of the various steps  involved in the \karmma forward model. Here, SHT refers to the Spherical harmonic transform, KS refers to Kaiser-Squires. Note that each of the steps illustrated above are differentiable, thus allowing us to compute the gradient with respect to the input variables -- a necessary feature for using HMC. The maximum $\ell$ value of the spherical harmonic modes and the {\sc healpix} map resolution of the maps is shown at each step. See section \ref{sec:formalism} for details. }
    \label{fig:fwd_model}
\end{figure*}

\begin{figure*}
    \centering
    \includegraphics[width=\linewidth]{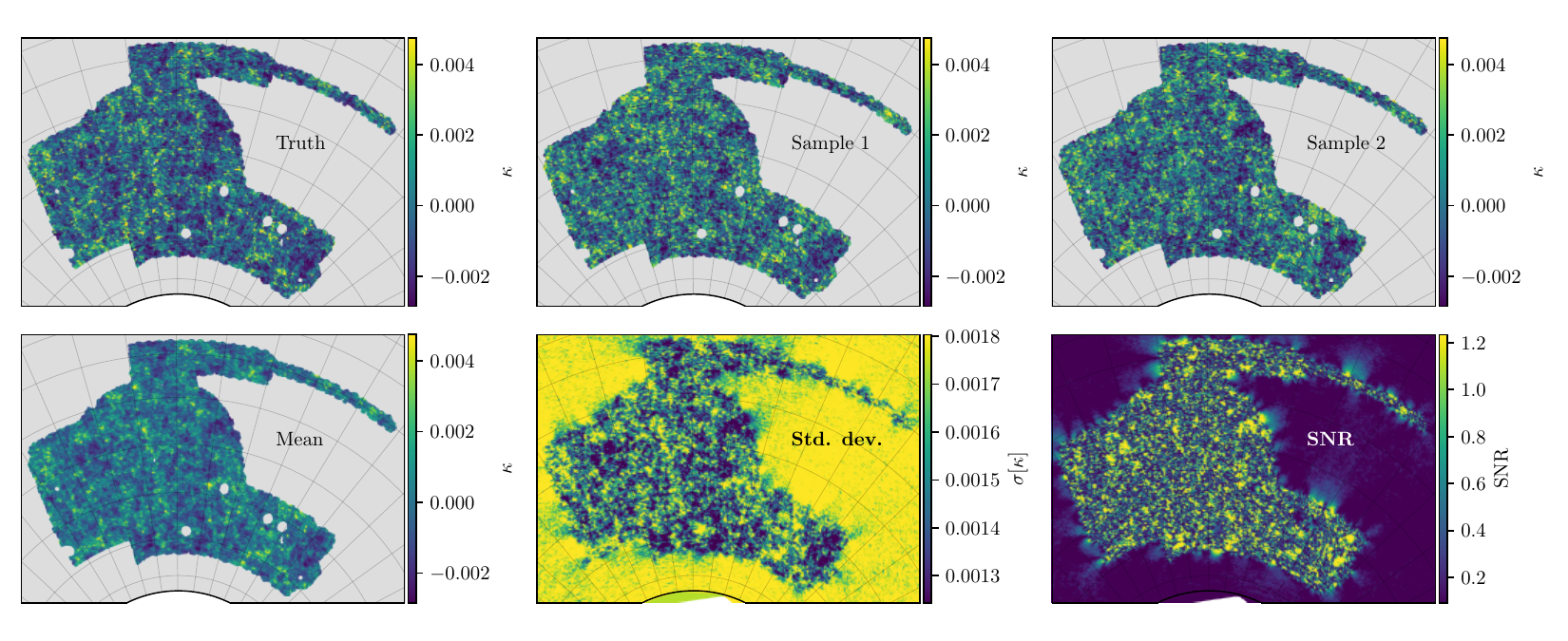}
    \caption{An illustration of the properties of the \karmma posteriors for the first tomographic redshift bin in our simulations. The input convergence map used to generate our mock observations is shown in the top left panel.  
    The centre and right panels on the top row shows two randomly chosen {\sc karmma} map samples. The mean map and the corresponding pixel-by-pixel RMS uncertainty are shown in the bottom left and bottom centre panels, respectively. The bottom right panel shows the signal to noise ratio computed from the sample mean and standard deviation.  
    }
    \label{fig:map_comparison_bin1}
\end{figure*}

\section{Mass map inference with \texttt{KaRMMa}}\label{sec:formalism}

\karmma\ models the convergence field as realizations of a lognormal random field.  The observed shears are then forward-modeled from the resulting convergence fields.  We determine the posterior distribution of convergence maps by sampling the posterior using Hamiltonian Monte Carlo (HMC) methods.  In this way, rather than providing a single ``best-fit'' convergence, we recover the full posterior distribution of convergence maps consistent with the input shear data.  These posterior quantify the uncertainty in the reconstructed mass maps.  We describe this algorithm in further detail below.  

Our lognormal assumption states that the convergence field $\kappa$ is a non-linear transform of a Gaussian random field $y$ where 
\begin{equation}\label{eqn:lognormal}
    \kappa = e^{y} - \lambda.
\end{equation} 
The parameter $\lambda$ is called the shift parameter. At the resolutions of this study ($N_{\rm side}=256$, or $\approx 13\ {\rm arcmin}$), a multivariate lognormal model provides a good description of the convergence field \citep{Xavier2016, Clerkin2017}.

\begin{figure*}
    \centering
    \includegraphics[width=\linewidth]{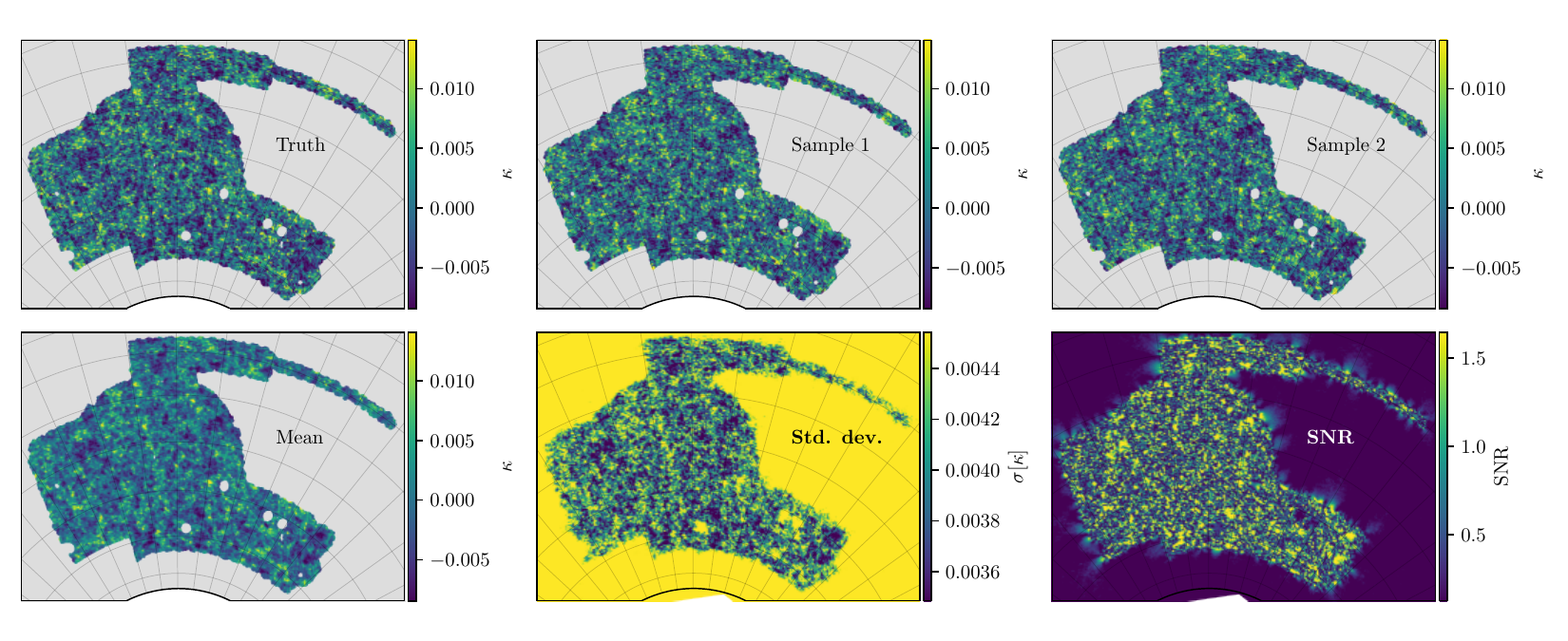}
    \caption{Same as Figure \ref{fig:map_comparison_bin1}, but for the fourth tomographic bin. Compared to Figure \ref{fig:map_comparison_bin1}, the structures are produced with a higher SNR in the higher redshift bins.}
    \label{fig:map_comparison_bin4}
\end{figure*}

Here, we use the lognormal formalism developed in \cite{Fiedorowicz2022, Boruah2022}.
Specifically, we sample the spherical harmonics (or Fourier wave modes) of the $y$ maps. Furthermore, following \cite{Boruah2022} we introduce new reparameterized variables, $\mvec{x}$, defined via 
\begin{equation}
    y_{lm} = \mmat{L}(l) x_{lm},
\end{equation}
where, $y_{lm}$ is the spherical harmonics of the $y$ variable defined in equation \eqref{eqn:lognormal} and $\mmat{L}(l)$ is the Cholesky decomposition of a $n_{\text{bin}} \times n_{\text{bin}}$ matrix constructed from the power spectrum $C(l)$ at a wave mode $l$ in various tomographic bins ($n_{\text{bin}}$ being the number of tomographic bins).\footnote{Note that the definition of the unit variance variables differ from the definition in \cite{Boruah2022} where the $x_{lm}$ variables were related to the $y_{lm}$ through the eigenvalue decomposition of $C(l)$. } This reparameterization de-correlates the variables in different redshift bins such that all the elements of $\mvec{x}$ have unit variance and no cross-correlation under the prior.\footnote{Note that posterior samples of $x$ can be correlated due to the likelihood.} As shown in \cite{Boruah2022}, this redefinition makes the sampling of mass maps more efficient. The power spectrum and the shift parameters are computed at the T17 cosmological parameters using their simulations. 

Following \cite{Fiedorowicz2022a}, we filter our maps at $\ell_{\text{max}} = 2 N_{\text{side}}$ to avoid aliasing. Note that the low-pass filter is applied on the $\kappa$ field and not on the $y$ (or $x$) fields. Also note that because of the filtering, the resulting $\kappa$ field is no longer a lognormal field.

Once we have the convergence field, the shear field is constructed using the Kaiser-Squires (KS) relation. We perform our inference on the sphere, thus requiring the curved sky version of the KS relation; the spherical harmonic coefficients of the shear field are related to the spherical harmonic coefficients of the $\kappa$ field via \citep{Wallis2022}
\begin{equation}
    \tilde{\gamma}_{lm} = \sqrt{\frac{(l-1)(l+2)}{l(l+1)}}\kappa_{lm}.
\end{equation}
The shear field is then obtained as a sum over the spin-2 spherical harmonics, 
\begin{equation}
    \gamma(\mvec{\theta}) = \sum_{lm} \tilde{\gamma}_{lm}~{}_2 Y_{lm}(\mvec{\theta}).
\end{equation}
Since the convergence field is low-pass filtered, the shear field is also low-pass filtered. By contrast, the observed shear field is \it not \rm filtered, which leads to some amount of model mis-specification.  However, we have found that given the resolution of our maps, this slight model mis-specification is easily preferable to the aliasing that occurs when we do not filter the field \cite{Fiedorowicz2022a}. 

Finally, we use pixelized shear maps to perform the inference, so we must account for the pixel window function in our forward model.  The pixelization operation is implemented in the harmonic space by multiplying each spherical harmonic mode by the spherical transform of the pixel window function,  
\begin{equation}
    \tilde{\gamma}_{lm} \rightarrow W^{N_{\text{side}}}_p(l)\tilde{\gamma}_{lm},
\end{equation}
where, $W^{N_{\text{side}}}_p$ is the {\sc healpix} pixel window function at resolution of $N_{\text{side}}$.  While in principle the pixel window function for each {\sc healpix} pixel is different, an average pixel window function can be computed using the {\sc healpy} function \texttt{healpy.sphtfunc.pixwin}. The need for including the pixel window function in the forward model has previously been noted in \cite{Fiedorowicz2022, Zhou2023}.

Given a $\kappa$ map, the observed shear map is modeled as a noisy realization of the true shear map, where the noise is set by the shape noise of the survey.  In particular, the noise is uncorrelated from pixel to pixel, so that
\begin{equation}\label{eqn:likelihood}
    \log P(\gamma_{\text{obs}}|\mvec{x}) = \frac{\chi^2}{2} = \sum_{i=1}^{N_{\text{pix}}} \frac{[\gamma^i(\mvec{x}) - \gamma^i_{\text{obs}}]^2}{2\sigma^2_{\eps,i}}.
\end{equation}
The summation is over all the pixels within the survey mask. 

Given an `observed' shear map, the posterior of $\mvec{x}$ is simply
\begin{equation}\label{eqn:posterior}
    P(\mvec{x}|\gamma_{\text{obs}}) \propto P(\gamma_{\text{obs }}|\mvec{x})P(\mvec{x}).
\end{equation}
Because $\mvec{x}$ are unit variance variables, the prior, $P(\mvec{x})$ is a unit variance normal distribution. 

The various steps involved in our forward model are illustrated in Figure \ref{fig:fwd_model}. Note that all the intermediate steps are differentiable allowing us to take the gradient of the posterior with respect to the variables $\mvec{x}$, as is required to sample using HMC. We implement our forward model in the python package {\sc pytorch} \citep{pytorch} and use the probabilistic programming language {\sc pyro} \citep{bingham2019pyro} to sample the mass-maps. We use the No U-Turn Sampler \citep[NUTS, ][]{Hoffman2011} to sample $\mvec{x}$ (or equivalently $\mvec{\kappa}$). 

We emphasize that in the current implementation, the transformation from the $x$ variables to convergence are conditioned on the power spectrum and shift parameter characterizing the lognormal prior.  Both of these are cosmology dependent, and therefore one must specify a cosmological model to sample from these maps.  One can, in principle, simultaneously sample the cosmological parameters characterizing the prior as part of the inference process as described in \cite{Boruah2022}.  In this way, \karmma can produce joint cosmology and mass map posteriors.  However, we have not yet characterized the shift parameter $\lambda$ as a function of cosmology, nor have we validated the resulting cosmological inference pipeline.  For this reason, we postpone this simultaneous inference process to future work.

%% file: sections/sim_test.tex

\begin{figure}
    \centering
    \includegraphics[width=\linewidth]{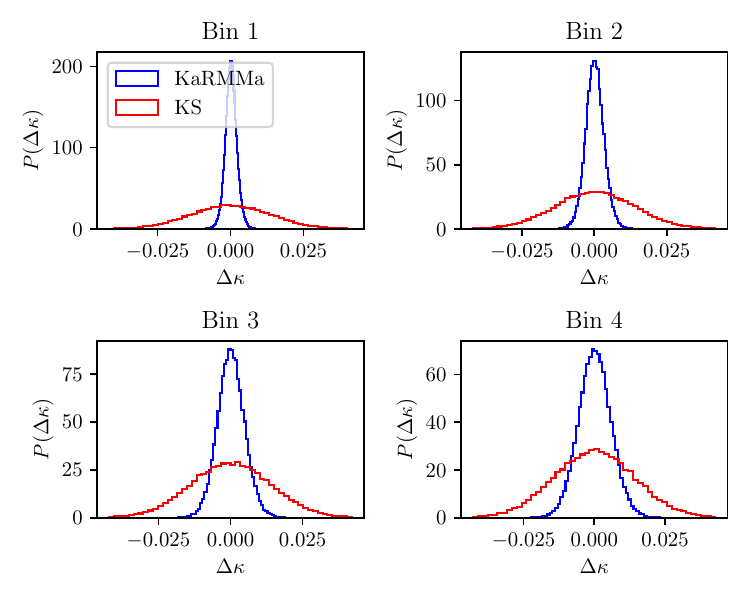}
    \caption{Comparison of the 
    residual errors in a randomly chose \karmma mass map sample ({\it blue}) and Kaiser-Squires mass map ({\it red}) created with the same data. The different panels show the results in different tomographic bins. As we can see from the figure, the residuals of the \karmma mass maps are much lower than that of the Kaiser-Squires mass maps, showing the better quality of the \karmma mass maps. The improvement is especially noteworthy for the lowest redshift bin, where the signal-to-noise is the lowest.
    }
    \label{fig:residual_hist}
\end{figure}

\begin{figure*}
    \centering
    \includegraphics[width=\linewidth]{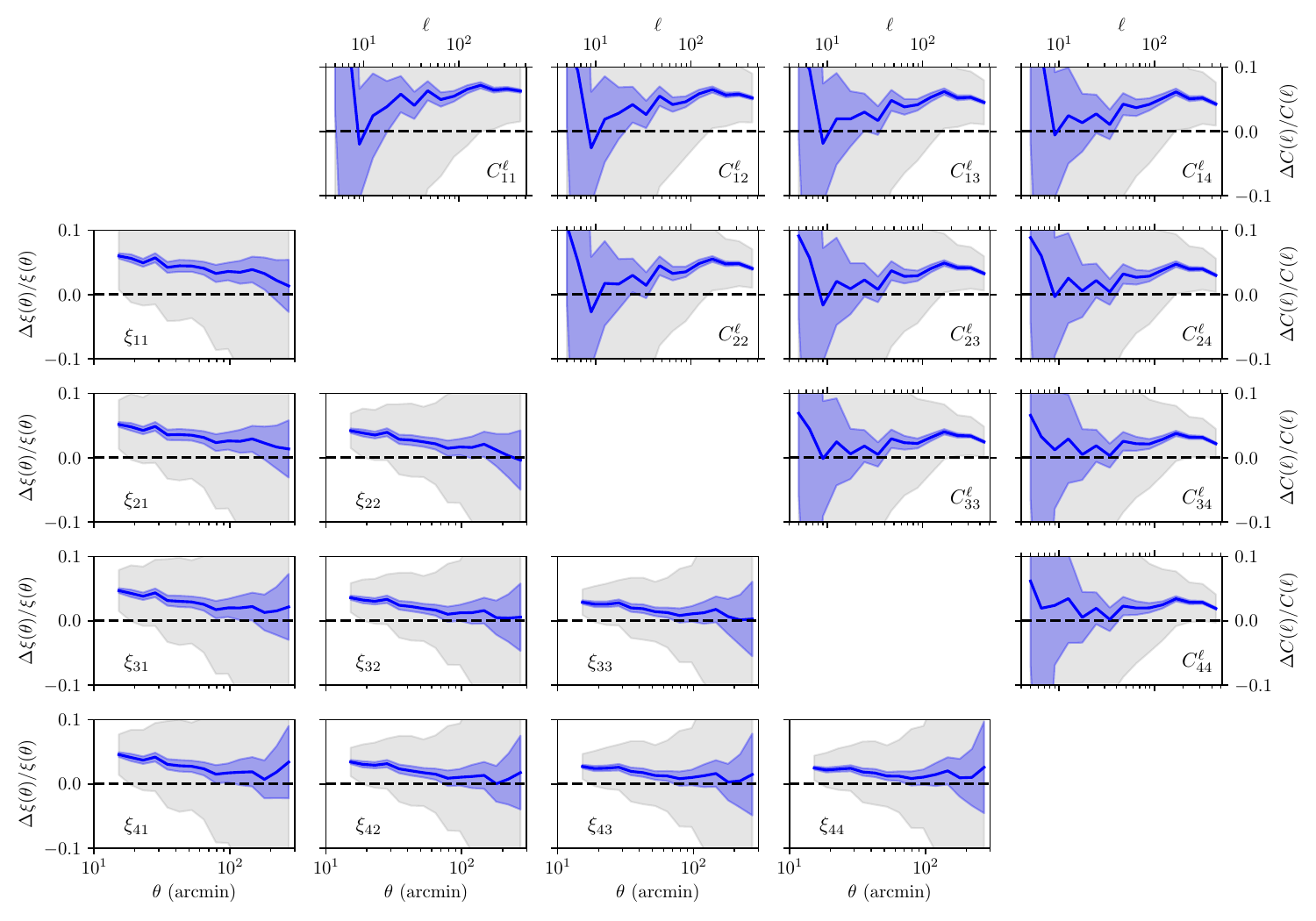}
    \caption{Fractional error between the two-point functions of the input simulation maps and those estimated from the \karmma\ posteriors.  The bottom-left triangles shows the correlation functions between the different tomographic bins, while the upper-right triangles shows the corresponding power spectra computed using pseudo-$C(l)$s.  Each panel corresponds to a different cross-bin combination, as labeled. The blue lines show the difference in the map statistics averaged across the full 108 simulated maps.  The grey bands shows the mean error in one map as estimated using the \karmma\ posteriors, while the blue band shows the error on the mean estimated using all 108 simulations. The difference between the simulated maps and our posteriors increases with decreasing redshift and smaller scales, and is typically $\lesssim 5\%$.}
    \label{fig:corr_cl} 
\end{figure*}

\begin{figure*}
    \centering
    \includegraphics[width=\linewidth]{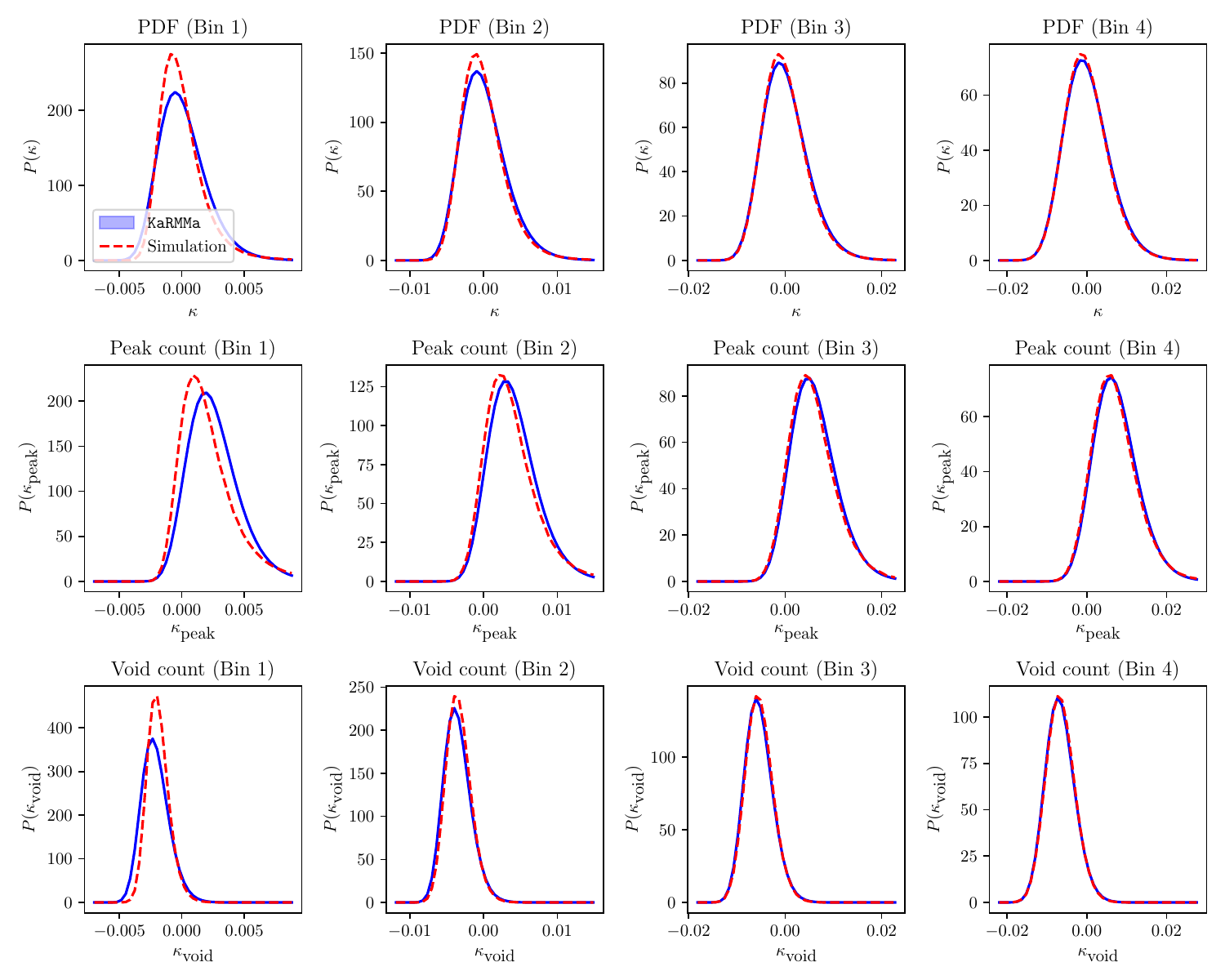}
    \caption{Comparison of non-Gaussian statistics of the \karmma maps {\it (blue)} and of the mock simulations {\it (red dashed)}. The different rows corresponds to the three non-Gaussian statistics computed here -- the PDF of $\kappa$ values ({\it top panels}), the peak counts ({\it middle panels}) and the void counts ({\it bottom panels}). The different columns show these statistics in the $4$ tomographic bins. As we can see, all these non-Gaussian summary statistics are well-approximated by \karmma at high redshift, whereas at low-redshift, these summary statistics are significantly biased due to the breakdown of the lognormal model.}
    \label{fig:ng_stats}
\end{figure*}

\section{Validating \texttt{KaRMMa} with mock simulations}\label{sec:sim_test}

We test the performance of \karmma using the simulations presented in section \ref{ssec:sims}. The multivariate lognormal prior requires the power spectrum and the shift parameters as its input. We measure these quantities in the simulations, averaging over all 108 simulations, and use these as the inputs to the lognormal prior.\footnote{We remind the reader that the use of the simulated power spectrum is necessary due to the small differences between the power spectrum in the simulations and theory power spectra.  When applied to data, we replace the power spectrum by the appropriate theory power spectrum for our fiducial cosmology.} We then run {\sc karmma} on all $108$ mock data sets created from the T17 data. This allows us to check how well the lognormal forward model performs on mock weak lensing data from $N$-body simulations. Each \karmma run requires $\sim \mathcal{O}(50-100)$ CPU core hours to produce $100$ independent samples on AMD Zen2 processors. 

Figures~\ref{fig:map_comparison_bin1} and \ref{fig:map_comparison_bin4} illustrate the properties of the \karmma\ posteriors when run on our simulated data set. The figures correspond to the mass maps from tomographic redshift bins 1 and 4 respectively. The input mass map used to generate the simulation is shown on the top left panel. The centre and right panels on the top row show two randomly chosen samples for our posterior distribution of maps. The mean and standard deviation of the posteriors are shown in the bottom left and bottom centre panels respectively. 
The bottom right panels shows the signal-to-noise ratio defined as,
\begin{equation}
    \text{SNR} = \frac{\text{Mean} (\kappa)}{\sigma[\kappa]}.
\end{equation}
By comparing the input convergence map to the \karmma posterior and mean map, we can see that \karmma recovers the dark matter distribution of the input map. We also see that the higher redshift bins have a higher signal-to-noise ratio, as expected.

Next we assess the quality of the reconstructed mass maps by computing the pixel-by-pixel errors in the output mass maps. The error between the convergence value in the $i$-th pixel of a reconstructed \karmma sample (labelled $s$) and the simulation is
\begin{equation}
    \Delta \kappa_i = \kappa^i_{\text{KARMMA},s} - \kappa^i_{\text{sim}}.
\end{equation}
In Figure \ref{fig:residual_hist}, we compare the histogram of the convergence residuals for two different maps: 1) a randomly chosen map from our posterior sample; and 2) a traditional Kaiser--Squires mass map.  Both maps are generated using the same input shear map.  We find the errors in the \karmma\ mass maps are always much lower than those in Kaiser--Squires maps. 
The standard deviation of the {\sc karmma} residuals are $85\%$, $77\%$, $66\%$ and $60\%$ lower than the standard deviations of the KS residuals in the $4$ tomographic bins respectively.
The improvement is especially noticeable for the lowest redshift bin, where the signal-to-noise is the lowest. As shown in \cite{Boruah2022}, adequately accounting for the covariance between tomographic bins dramatically improves the signal-to-noise of the $\kappa$ reconstruction at low redshift by allowing us to extract information about low redshift structures contained in the high redshift tomographic bins.

As we now demonstrate, one of the principal advantages of \karmma\ is that the posterior maps faithfully reproduce the statistical properties of the input convergence field.  Figure \ref{fig:corr_cl} shows the difference between the mean two-point function in our posterior maps and that of the input convergence map in both real and harmonic space.  For the latter, we compute pseudo-$C(l)$s using the publicly available code {\sc namaster} \citep{Alonso2019}. The blue line shows the mean difference, averaged across all 108 simulations.  The grey bands show the mean error in the posterior for one simulation, while the blue bands show the error on the mean.  The difference in the two point functions between the input simulated maps and our reconstructed maps is $\lesssim 5\%$, with the difference increasing with decreasing redshift.  These biases arise because of the failure of the lognormal model to accurately model non-linear structure formation. In particular, Appendix~\ref{app:ln_tests} demonstrates that we do not observe any such biases when we test \karmma\ using log-normal convergence maps as input.

Apart from the 2-point function, the lognormal model also captures many non-Gaussian features of the convergence field. Figure \ref{fig:ng_stats} compares three such statistics, namely the PDF of $\kappa$ values in each pixel (i.e. the one-point function), and the peak and void counts of the $\kappa$ maps. The peaks (voids) of the map are defined as the number of pixels which have the highest (lowest) $\kappa$ value among all its neighbors. These statistics are well-known to be sensitive to the non-Gaussian information of the convergence field \citep{Liu2015a, Davies2021}. We see that the non-Gaussian statistics of the \karmma maps are close to those in the simulations, particularly for the high redshift bins. However, the differences between the simulated maps and our posteriors increases with decreasing redshift, being particularly large for the lowest redshift bin. This is not surprising, as it is at this redshift that non-linear structure formation is most important.  We note that because the signal-to-noise of the convergence in an individual pixel is low ($S/N \lesssim 1$), the one point of the posteriors is nearly identical to that of the prior.

%% file: sections/results.tex
\section{Bayesian mass maps with DES-Y3 data}\label{sec:results}

\begin{figure*}
    \centering
    \includegraphics[width=\linewidth]{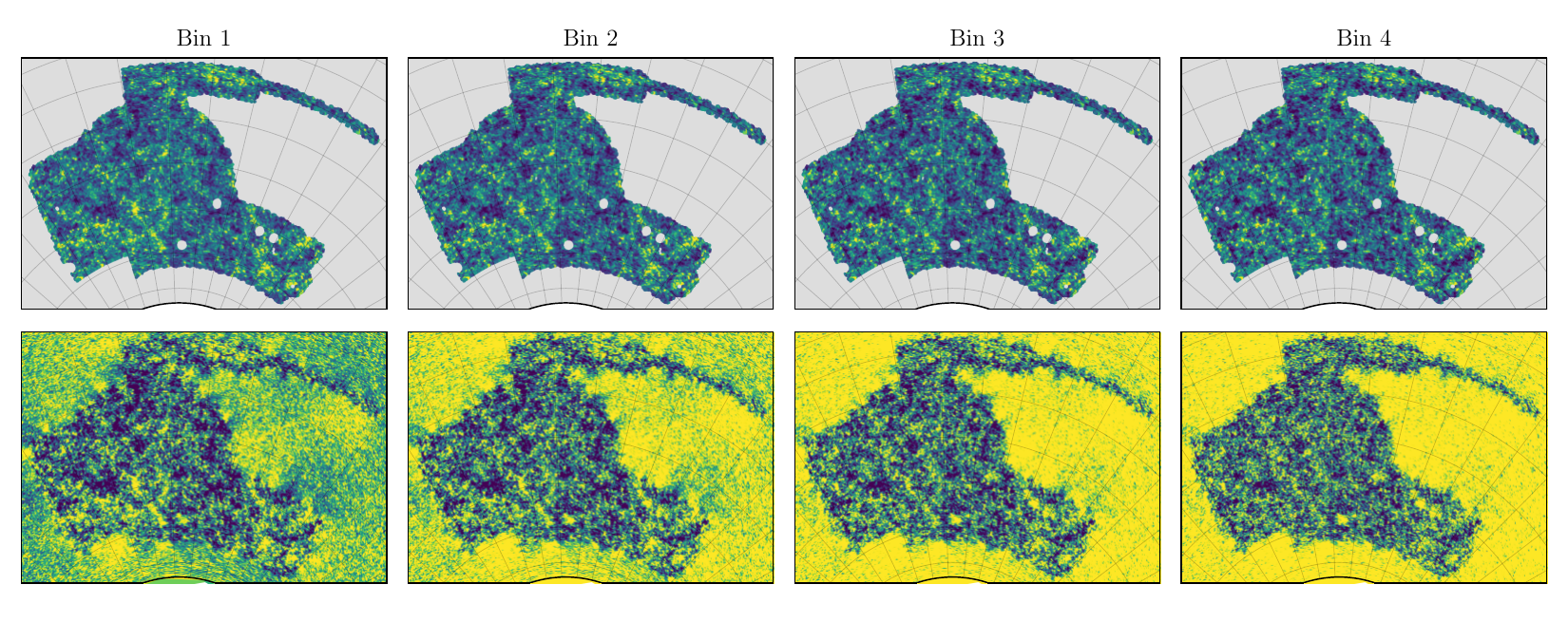}
    \caption{The mean ({\it top}) and standard deviation ({\it bottom}) of the \karmma posterior maps for DES-Y3 data in each of the 4 tomographic bins, as labeled.}
    \label{fig:desy3_mean_std}
\end{figure*}

We run \karmma on pixelized shear maps produced from the DES-Y3 shape catalogs as detailed in section \ref{ssec:desy3_data}. The input $C(l)$ for this run is computed for the T17 cosmological parameters using {\sc pyccl} \cite{pyccl}. We use the shift parameters computed from the T17 simulations for these \karmma runs. We produce a chain with $500$ independent samples. The maps showing the mean and standard deviation of the posteriors in each of the four tomographic bins is shown in Figure~\ref{fig:desy3_mean_std}.The figure highlights the fact that \karmma\ produces full posterior distributions of the mass maps, which is particularly important if these maps are to be used for cross-correlation studies \citep[e.g.][]{Nguyen2020}.

\begin{figure*}
    \centering
    \includegraphics[width=\linewidth]{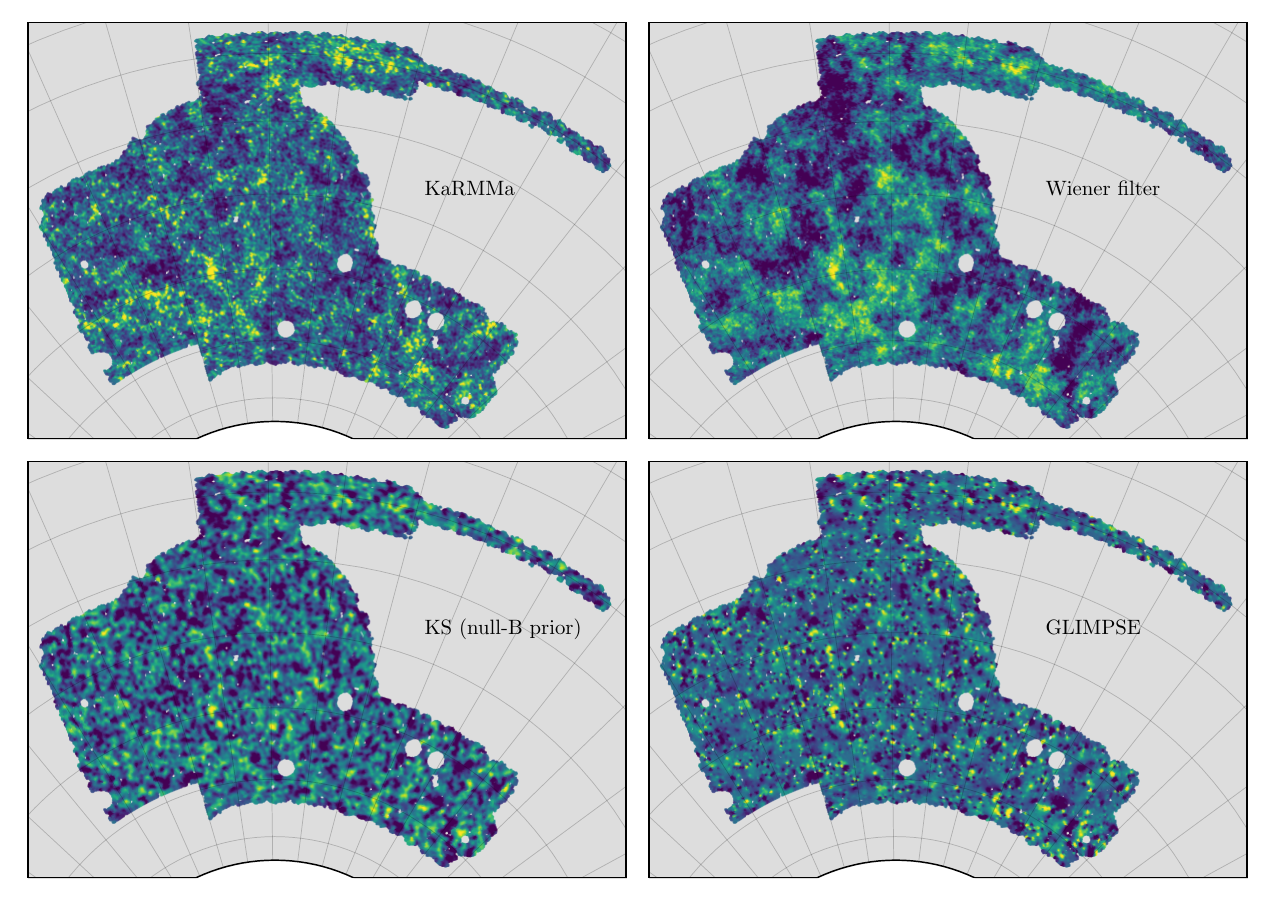}
    \caption{Comparison of the mass map for the lowest tomographic redshift bin in the DES-Y3 data.  Each panel corresponds to a different algorithm: \karmma mean map ({\it top left}), Wiener filtering ({\it top right}), Kaiser-Squires with null B-mode prior ({\it bottom left}), and {\sc glimpse} ({\it bottom right}). The fact that \karmma adequately models the covariance between tomographic redshift bins enable us to robustly recover structure on scales significantly smaller than those resolved in other algorithms.}
    \label{fig:desy3_map_comparison}
\end{figure*}

Figure \ref{fig:desy3_map_comparison} compares the mean of our posterior maps for the lowest tomographic redshift bin to the publicly available mass maps released by the DES collaboration \cite{DESY3_massmapping}.  Specifically, we compare our mean mass map to their Null B-mode prior Kaiser-Squires method, Wiener filter method, and the wavelet transform based {\sc glimpse} method \citep{Lanusse2016}. We find that \karmma can resolve smaller structures than any of the other methods. This improvement is driven by the fact that \karmma self-consistently accounts for the covariance across all tomographic redshift bins, which, as noted earlier, allows us to extract information on the low-redshift density field from the high-redshift shear data \citep{Boruah2022}. 

\begin{figure*}
    \centering
    \includegraphics[width=\linewidth]{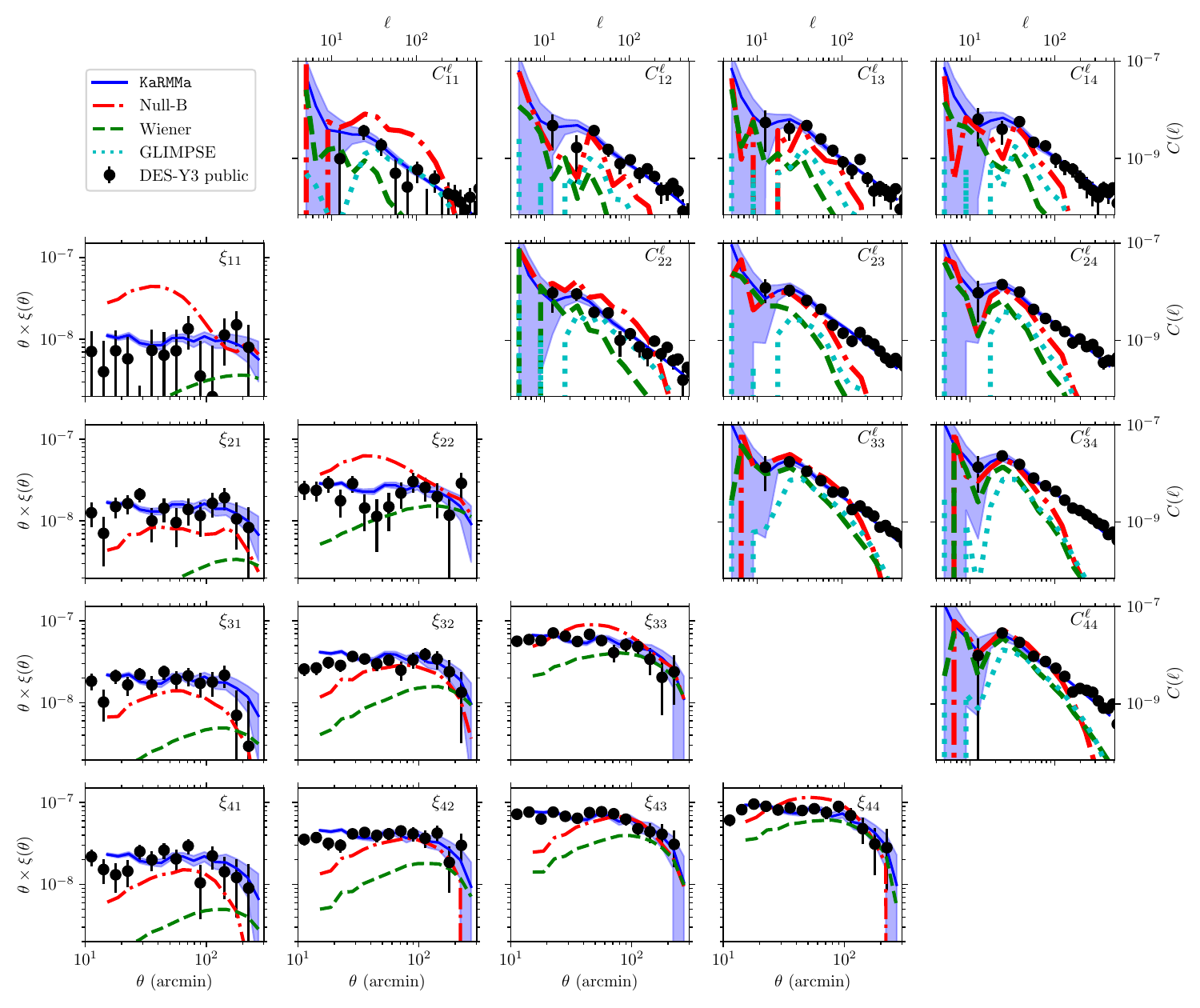}
    \caption{Comparison of the correlation function ({\it bottom left}) and the pseudo-$C(\ell)$ ({\it top right}) of mass maps created from DES-Y3 data with the same data vectors computed directly from the DES-Y3 shape catalogue ({\it black}). The blue shaded region shows the $95\%$ confidence interval of the summary statistics of {\sc karmma} posteriors. The correlation function and pseudo-$C(\ell)$'s of DES-Y3 mass maps created with different methods, namely Null-B prior KS ({\it red dash dotted}), Wiener filter ({\it green dashed}), {\sc glimpse} ({\it cyan dotted}), are also shown in the figure. As we can see from the figure, none of the mass mapping methods except {\sc karmma} reproduce the correlation functions of the DES-Y3 weak lensing data. }
    \label{fig:desy3_corr_cl_comparison}
\end{figure*}

In Figure \ref{fig:desy3_corr_cl_comparison}, we compare the correlation function and the pseudo $C(l)$ of the DES-Y3 mass maps reconstructed using \karmma and the same publicly available mass maps used in Figure \ref{fig:desy3_map_comparison}. In that figure, we also show the correlation functions and the pseudo-$C(l)$s measured by the DES collaboration with black dots.\footnote{DES-Y3 data vectors are available at \url{https://des.ncsa.illinois.edu/releases/y3a2/Y3key-products} and the pseudo-$C(l)$ measurements are taken from \cite{Doux2022}.} As we can see from the figure, barring \karmma, none of the other methods recover the expected 2-point functions in the mass maps. This is unsurprising.  Indeed, the fact that best point-estimates of the convergence map are expected to be biased further highlights the importance of being able to properly sample the posterior distribution of the maps.

\begin{figure*}
    \centering
    \includegraphics[width=\linewidth]{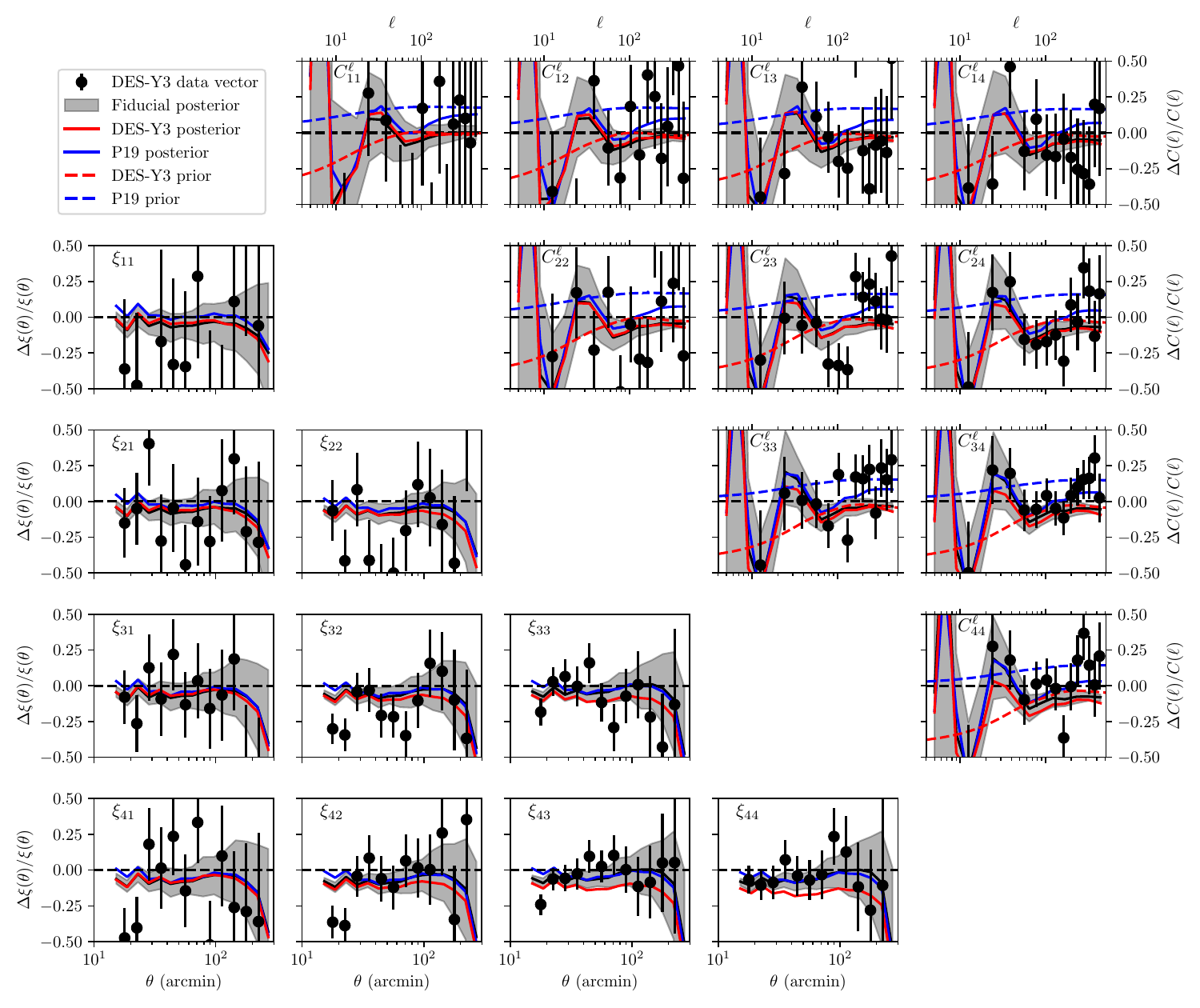}
    \caption{The fractional difference in the recovered correlation function and pseudo $C(\ell)$'s for {\sc karmma} runs with different cosmological parameters with respect to the theory predictions from T17 cosmology. The grey shaded region shows the $95\%$ confidence interval for the {\sc karmma} run with the fiducial cosmological parameters of T17. The blue (red) solid line shows the mean of the {\sc karmma} posterior with Planck (DES-Y3) cosmological parameters. The blue (red) dashed lines in the $C(\ell)$ plot shows the input $C(\ell)$ prior for the corresponding cosmological parameters. As we can see from the figure, the $C(\ell)$'s on large scales are not substantially impacted by the input cosmological parameters. However, the small-scale power depends on the input cosmological parameters.}
    \label{fig:desy3_delta_corr_cl}
\end{figure*}

We run 2 additional {\sc karmma} analysis with different priors to test the sensitivity of the recovered mass maps to the input cosmological parameters. We use the best fit cosmological parameters from Planck Collaboration cosmological analysis \cite{Planck2020} and DES-Y3 3$\times$2 pt analysis \cite{DESY3_3x2pt_2022}. For each of these cosmologies, we recompute the power spectrum and shift parameters. The shift parameter at different cosmologies is computed by rescaling the shift parameter in the T17 simulations with the ratio of the shift parameter predicted using {\sc cosmomentum} \citep{Friedrich2020}.  That is, we only rely on analytic predictions for the shift parameter for the purposes of rescaling the shift parameter from our fiducial cosmology to a new cosmology. The results of these runs are shown in Figure \ref{fig:desy3_delta_corr_cl}, where we plot the fractional difference of the recovered correlation function and pseudo-$C(\ell)$'s for the \karmma runs with different cosmological parameters with respect to the T17 2-point functions. As we can see from the figure, the power spectrum on large scales is largely insensitive to the input prior. On large scales, the signal-to noise is high and therefore the large-scale modes are determined by the likelihood. By contrast, the signal-to-noise ratio on small scales is low, which renders the posterior at small scales sensitive to the adopted prior. 

\begin{figure}
    \centering
    \includegraphics[width=\linewidth]{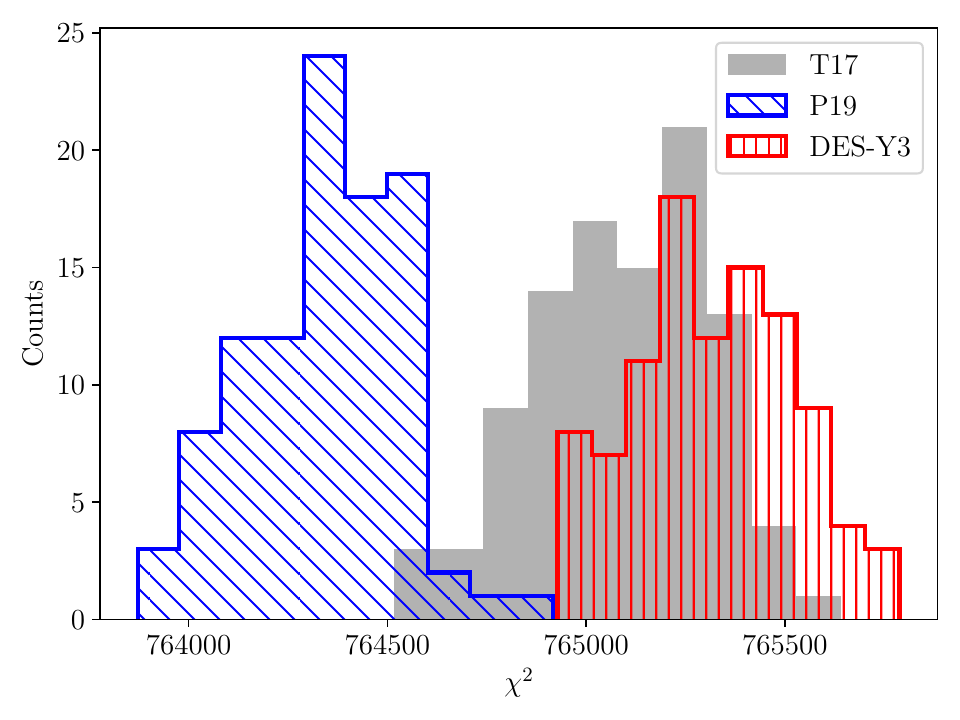}
    \caption{$\chi^2$ distribution from the \karmma posteriors with different cosmological parameters. The posteriors with DES-Y3 cosmological parameters ({\it red vertical lines}) have the highest $\chi^2$, followed by the T17 ({\it gray histograms}) and P19 ({\it blue diagonal lines}). The fact that these distributions are clearly differentiated suggests a field-level cosmological analysis of the DES shear maps may be able to distinguish between the Planck and DES-Y3 cosmologies.}
    \label{fig:chi_sq_dist_desy3}
\end{figure}

In Figure \ref{fig:chi_sq_dist_desy3} we compare the distribution of $\chi^2$ values computed using equation \eqref{eqn:likelihood} for each map in the \karmma posterior. Interestingly, we see that the \karmma maps produced using a Planck cosmology provide a better fit for the DES-Y3 data than the DES-Y3 or T17 cosmological parameters. However, we caution against interpreting this result as evidence for a Planck cosmology since we did not account for observational systematics in our analysis.  Rather, we want to highlight the difference in the likelihood between the Planck and DES-Y3 posteriors suggests that a field-level analysis of the DES-Y3 data may have the potential to distinguish between these different cosmologies.

%% file: sections/conclusion.tex
\section{Conclusion}\label{sec:conclusion}

The \karmma\ algorithm \citep{Fiedorowicz2022, Fiedorowicz2022a} models the convergence field as a lognormal random field.  By doing so, \karmma can perform a forward-modelled Bayesian reconstruction of the density field of the Universe as constrained by cosmic shear data.  In this paper, we have extended the \karmma framework so that it can simultaneously forward model multiple tomographic redshift bins while fully accounting for their expected covariance.  This enhancement significantly improves the signal-to-noise of the reconstructed maps, particuarly in the low signal-to-noise regime (See section \ref{sec:sim_test} and \cite{Boruah2022}). We validated our method on simulated cosmic shear data generated using $N$-body simulations, and demonstrated that the \karmma posteriors accurately reproduce a variety of Gaussian and non-Gaussian statistics of the input simulated maps (see Figures~\ref{fig:corr_cl} and \ref{fig:ng_stats}).  However, we do find evidence of small $\lesssim 5\%-10\%$ biases which increase with decreasing redshifts.  These biases arise due to the failure of the lognormal model to correctly capture non-linear structure formation.  These results suggest that further improvements require improving the field-level prior assumed for the convergence field. Some options for such an improvement include analytic extensions such as the double log model of \cite{Porth2023} or using generative artificial intelligence models \citep{Fiedorowicz2022ml, Yiu2022, Dai2023, Shirasaki2023}.

Following our simulation tests, we applied \karmma\ to DES-Y3 weak lensing data to produce the first Bayesian forward-modelled tomographic mass maps from Stage-III weak lensing data on a sphere.  These maps are also the first to self-consistently account for the covariance between tomographic redshift bins as part of the mass map reconstruction algorithm, which in turn improved the signal-to-noise of the resulting maps, particularly at low redshifts. We show that these maps have the correct theoretically expected statistical properties such as the $2$-point function. We make these maps publicly available at \url{https://zenodo.org/records/10672062}.

An important limitation in our work is the relatively coarse angular resolution of our lensing maps ($\approx 13\ {\rm arcmin}$).  Since the information on small-scales is highly non-Gaussian, one might expect that field-level inference will become even more useful at higher resolution.  However, this leads to increased computational demands due to: {\it i)} spherical harmonic transforms at a higher resolution; and {\it ii)} a larger parameter space. Significantly improving resolution over that achieved in this work will require improved computational spherical harmonics code and faster sampling methods. Improved spherical harmonic transform codes have been produced using GPU acceleration \citep[e.g,][]{Tian2022} or spherical Fourier neural operators \citep{Bonev2023}. Improvement in sampling can be achieved using improved sampling methods such as microcanonical Hamiltonian Monte-Carlo methods \citep{Robnik2022, Bayer2023a}. 

Despite this limitation, the inclusion of tomographic mass mapping on a sphere represents an important step towards enabling a full field-level cosmological analysis of cosmic shear data. In addition to varying cosmology, our future work will focus on incorporating weak lensing systematics as part of the inference process.  The field-level approach is particularly well-suited for incorporating some of these systematics. For example, it has been shown that the non-Gaussian clustering of galaxies can improve the inference of photometric redshifts \citep{Jasche2012, Tsaprazi2023}. Likewise, Bayesian photo-$z$ inference algorithms \citep{Leistedt2016, Leistedt2023, Sanchez2019, Alarcon2020} can naturally be extended to incorporate density field inference to further improve photometric redshift estimation.  Similarly, we will need to model intrinsic alignments at the field level in order to properly extract information using our forward modeling framework \citep{Tsaprazi2022}.  Fortunately, we see no reason why any of these challenges should be prohibitive, suggesting that field-based inference of cosmological parameters from cosmic shear data could soon be realized.

%% file: sections/appendix/ln_test.tex
\section{Tests with lognormal mocks}\label{app:ln_tests}

\begin{figure*}
    \centering
    \includegraphics[width=\linewidth]{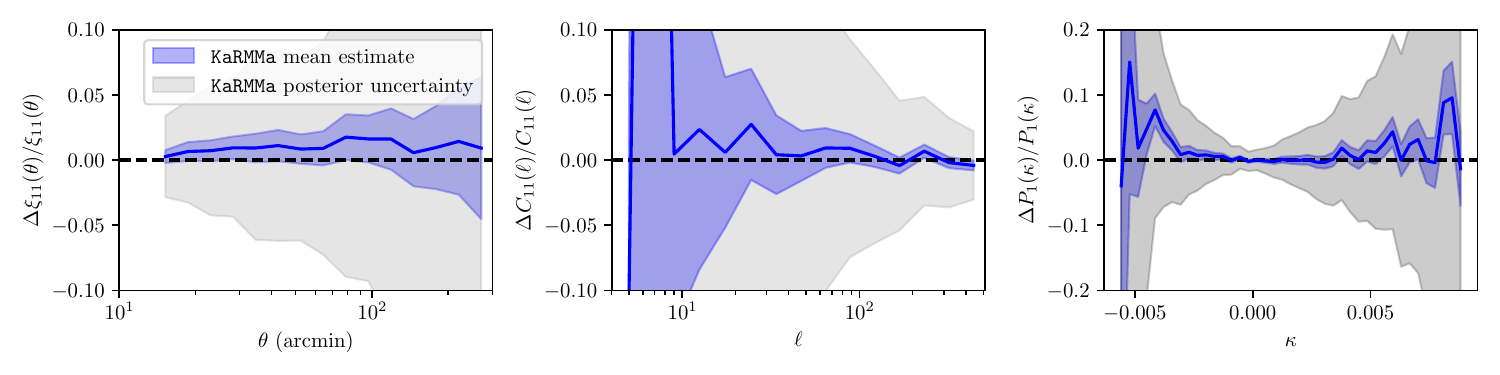}
    \caption{Fractional error in the correlation function ({\it  left}), pseudo-$C(\ell)$ ({\it centre}) and the 1-pt PDF ({\it right}) of the first redshift bin of the \karmma posteriors for runs on lognormal mocks. As we can see, we do not see the large biases in these summary statistics that we saw for the \karmma runs on $N$-body simulations in section \ref{sec:sim_test}. Thus, it shows that the biases seen in section \ref{sec:sim_test} and Figures \ref{fig:corr_cl}, \ref{fig:ng_stats} arises due to the model mis-specification due to the lognormal prior.}
    \label{fig:LN_test}
\end{figure*}

In Figure \ref{fig:LN_test}, we show the fractional error in the correlation function, pseudo-$C(\ell)$'s and the 1-pt PDF in the \karmma posterior maps on runs with lognormal mocks. For the ease of visualization, we only show the results for the first redshift bins, where we saw in section \ref{sec:sim_test} that the uncertainty was the largest. As we can see from the figure, we do not see the $\sim \mathcal{O}(5\%)$ bias in the 2-pt functions or the even larger bias in the 1-pt PDF. This demonstrates that the biases we see in our runs with $N$-body simulations arise because of model mis-specification due to the assumed lognormal prior. 